\documentclass[12pt]{article}
\usepackage{epsfig}
\def\mrm{\mathrm}

\newcommand{\epm} {\mbox{$\mathrm{e}^+ \mathrm{e}^-$}}

\newcommand{\qqb}{\mbox{$\mathrm{q\overline{q}}$}}

\newcommand{\lnu}{\mbox{$\ell\bar{\nu}_{\ell}$}}

\newcommand{\mW}{\mbox{$m_{\mathrm{W}}$}}
\newcommand{\mZ}{\mbox{$m_{\mathrm{Z}}$}}
\newcommand{\mtop}{\mbox{$m_{\mathrm{t}}$}}

\newcommand{\qq}         {\mbox{$\mathrm{q}\bar{\mathrm{q}}$}}

\newcommand{\Zgs}        {\mbox{$\mathrm{(Z/\gamma)}^{*}$}}

\newcommand{\WW}         {\mbox{$\mathrm{W}\mathrm{W}$}}
\newcommand{\pb}         {\mbox{$\mathrm{pb}^{-1}$}}

\def\mrm       {\mathrm}

\newcommand{\NPhys}     {Nucl.~Phys.}

\def\etal{\mbox{{\it et al.}}}

\newcommand{\NIMA}[3] {Nucl.\ Instr.\ Meth.\ \textbf{A#1} (#2) #3}
\newcommand{\CiP}[3]  {Comp.\ in Phys.\ \textbf{#1} (#2) #3}
\newcommand{\myWW}         {\mbox{$\mathrm{W}\mathrm{W}$}}
\newcommand{\qqln}{\mbox{$\qq\lnu$}}

\newcommand{\Wboson}{\mbox{$\mathrm{W}$}}
\newcommand{\Zboson}{\mbox{$\mathrm{Z}$}}
\newcommand{\Vboson}{\mbox{${V}$}}

\newcommand{\roots}{\mbox{$\sqrt{s}$}}

\newcommand{\WWqqln}{\mbox{$\myWW\rightarrow\qq\lnu$}}
\newcommand{\eeWW}{\mbox{${\rm e^+e^-\rightarrow} \WW$}}

\newcommand{\Vub} {\mbox{$|V_{\mathrm{ub}}|$}}
\newcommand{\Vcb} {\mbox{$|V_{\mathrm{cb}}|$}}
\def\CL{\mbox{{\rm CL}}}

\newcommand{\Wqq}{\mbox{$\Wboson\rightarrow\qq$}}
\newcommand{\Wlnu}{\mbox{$\Wboson\rightarrow\lnu$}}

\newcommand{\WpWm}{\mbox{$\mathrm{W}^+\mathrm{W}^-$}}
\newcommand{\epem}{\mbox{$\mathrm{e}^+\mathrm{e}^-$}}
\newcommand{\qqqq}{\mbox{$\qq\qq$}}

\newcommand{\myPLB}[3]  {Phys.\ Lett.\ \textbf{B#1} (#2) #3}
\newcommand{\myZPC}[3]  {Z.\ Phys.\ \textbf{C#1} (#2) #3}
\newcommand{\myEPC}[3]  {Eur.\ Phys.\ J.\ \textbf{C#1} (#2) #3}

\def\opalabbiendi{OPAL Collaboration, G. Abbiendi \etal}
\def\opalalexander{OPAL Collaboration, G.\ Alexander \etal}

\newcommand{\ra}        {\mbox{$\rightarrow$}}   % \to can be used as well
\begin{document}
\begin{titlepage}
\begin{center}
{\Large  EUROPEAN ORGANIZATION FOR NUCLEAR RESEARCH}
\end{center}
\bigskip\bigskip
\begin{flushright}
       CERN-EP-2001-066 \\
       20 September 2001
\end{flushright}
%\bigskip\bigskip\bigskip\bigskip\bigskip
\begin{center}
{\huge\bf\boldmath  Search for Single Top Quark \\
 Production at LEP2 \\}
\end{center}
\bigskip
\bigskip
\begin{center}{\LARGE The OPAL Collaboration}\end{center}
\bigskip\bigskip
\bigskip
\begin{center}{\large  Abstract}\end{center}
A search for single top quark production via flavour changing neutral
currents (FCNC) was performed with data collected by the OPAL
detector at the ${\rm e^+e^- }$ collider LEP. Approximately 600
pb$^{-1}$ of data collected at $\sqrt{s}$ = 189\,-\,209\,GeV were used
to search for the FCNC process ${\rm e^+ e^- \rightarrow tc(u)
\rightarrow bWc(u)}$. This analysis is sensitive to the leptonic
and the hadronic decay modes of the W boson.  No evidence for a
FCNC process is observed.  Upper limits at the 95\% confidence level on the
single top production cross-section as a function of the
centre-of-mass energy are derived.  Limits on the anomalous coupling
parameters $\kappa_{\gamma}$ and $\kappa_{\rm Z}$ are determined from
these results.
\mbox{ }\\
\bigskip\bigskip\bigskip
\begin{center} {\large (Submitted to Physics Letters B)}
\end{center}
\end{titlepage}
%
%-----------------------------------------------------------------------
%
\begin{center}{\Large        The OPAL Collaboration
}\end{center}\bigskip
\begin{center}{
%begin authorlist PLEASE DO NOT DELETE THIS COMMENT
G.\thinspace Abbiendi$^{  2}$,
C.\thinspace Ainsley$^{  5}$,
P.F.\thinspace {\AA}kesson$^{  3}$,
G.\thinspace Alexander$^{ 22}$,
J.\thinspace Allison$^{ 16}$,
G.\thinspace Anagnostou$^{  1}$,
K.J.\thinspace Anderson$^{  9}$,
S.\thinspace Arcelli$^{ 17}$,
S.\thinspace Asai$^{ 23}$,
D.\thinspace Axen$^{ 27}$
% ...
%end authorlist PLEASE DO NOT DELETE THIS COMMENT
}\end{center}\bigskip
\bigskip
%begin institutes
$^{  1}$School of Physics and Astronomy, University of Birmingham,
Birmingham B15 2TT, UK
\newline
$^{  2}$Dipartimento di Fisica dell' Universit\`a di Bologna and INFN,
I-40126 Bologna, Italy
\newline
$^{  3}$Physikalisches Institut, Universit\"at Bonn,
D-53115 Bonn, Germany
\newline
% ...
$^{  5}$Cavendish Laboratory, Cambridge CB3 0HE, UK
\newline
% ...
$^{  9}$Enrico Fermi Institute and Department of Physics,
University of Chicago, Chicago IL 60637, USA
\newline
% ...
$^{ 16}$Department of Physics, Schuster Laboratory, The University,
Manchester M13 9PL, UK
\newline
$^{ 17}$Department of Physics, University of Maryland,
College Park, MD 20742, USA
\newline
% ...
$^{ 22}$Department of Physics and Astronomy, Tel Aviv University,
Tel Aviv 69978, Israel
\newline
$^{ 23}$International Centre for Elementary Particle Physics and
Department of Physics, University of Tokyo, Tokyo 113-0033, and
Kobe University, Kobe 657-8501, Japan
\newline
% ...
$^{ 27}$University of British Columbia, Department of Physics,
Vancouver BC V6T 1Z1, Canada
\newline
% ...
Tel Aviv 69978, Israel.
%end notes

\newpage

\section{Introduction}
In the mid 1990's, the LEP collider at CERN entered a new phase of operation,
LEP2, with the first \epem\ collisions above the \WpWm\ threshold.
Between 1998 and 2000, with the installation of additional
super-conducting radio-frequency accelerating cavities, the
centre-of-mass energy of the LEP collider was further increased.  The
LEP2 data accumulated at centre-of-mass energies between 189\,GeV and
209\,GeV have opened up a new kinematic domain for particle searches.

The top quark mass was measured at the Tevatron collider to be
174.3 $\pm$ 5.1\,GeV$/c^{2}$~\cite{tev,pdg}. Due to this high mass, 
top quarks may only be singly produced at LEP2. Single
top quark production in the Standard Model (SM) process ${\rm e^+e^-
\rightarrow e^-\bar{\nu}_{e}t\bar{b}}$ has a cross-section of about
10$^{-4}$\,fb at LEP2 energies~\cite{top-lep200} and can not be seen
with the available luminosities. Another possible process for single top
quark production is the flavour changing neutral current (FCNC)
reaction\footnote{Throughout this paper, charge conjugate states are
implied.}:

\begin{equation}
\mrm{e^+e^-\rightarrow\bar{t}c(u)}.
\end{equation}
\label{eq:main}

Such FCNC are known to be absent at the tree level in the SM but can
naturally appear at the one-loop level due to CKM mixing which leads
to cross-sections of the order of $10^{-9}$\,fb at LEP2
energies~\cite{top-lep500}. Extensions of the SM such as
supersymmetry, exotic quarks, and multi-Higgs doublet models could
lead to an enhancement of such 
transitions~\cite{fritzsch,obrazt,han,aguila}. In this paper the search
for single top production via the FCNC reaction ${\rm
e^+e^-\rightarrow\bar{t}c(u)}$ is reported.

At the Tevatron, the CDF Collaboration performed a search for FCNC
in the top decays $\mrm{t\rightarrow \gamma\,c(u)}$ and
$\mrm{t\rightarrow Z\,c(u)}$ in $\mrm{p\bar{p}}$ collisions at a
centre-of-mass energy of 1.8\,TeV. They obtained upper limits at the
95\% confidence level (\CL) on the branching fractions~\cite{cdf}:
$\mrm{\mbox{Br}(t\rightarrow c\gamma)+\mbox{Br}(t\rightarrow
u\gamma)<3.2\%}$ and $\mrm{\mbox{Br}(t\rightarrow
cZ)+\mbox{Br}(t\rightarrow uZ)<33\%}$.

The FCNC reaction can be described with the parameters
$\kappa_{\gamma}$ and $\kappa_{\rm Z}$ which represent
the tree-level $\gamma$ and $\mathrm{Z}$ exchange contributions to
${\rm e^+e^- \to \bar{t}c(u)}$.  Thus, the Born-level cross-section 
for single top production in ${\rm e^{+}e^{-}}$ collisions for 
$\sqrt{s}>m_{\rm t}$ can be written as~\cite{obrazt}:
\begin{eqnarray}
 & & \sigma[{\rm e^+ e^- \to \bar{t} c(u)}] = \frac{\pi \alpha^2}{s}
\Biggl ( 1 - \frac{m^2_{\rm t}}{s} \Biggr )^2 \Biggl [
\kappa^2_{\gamma} e^2_q
\frac{s}{m^2_{\rm t}} \biggl ( 1 + \frac{2m^2_{\rm t}}{s}\biggr )
\nonumber \\ &+& \frac{ \kappa^2_{\rm Z} (1+a^2_w)(2+\frac{m^2_{\rm t}}{s})}
{4 \sin^4 2\vartheta_{\rm W} (1-\frac{m^2_{\rm Z}}{s})^2} + 3
\kappa_{\gamma} \kappa_{\rm Z}
\frac{ a_w e_q } {\sin^2 2\vartheta_{\rm W}
(1-\frac{m^2_{\rm Z}}{s})} \Biggr ],
\label{sintop:cross}
\end{eqnarray}
where $s$ is the centre-of-mass energy squared, $\alpha$ is the fine 
structure constant, $e_q=2/3$ and $m_{\rm t}$ are the charge and mass 
of the top quark, $m_{\rm Z}$ is the $\mathrm{Z}$ boson mass, and $a_w =
1-4\sin^2\vartheta_{\rm W}$ with $\vartheta_{\rm W}$ being the weak-mixing 
angle. The three terms in Equation~\ref{sintop:cross} correspond to
the contribution from annihilation via a photon, a $\mathrm{Z}$ boson,
and their interference.  Using the published limits of
CDF on FCNC, one can derive the following model-dependent
limits at 95\% \CL~\cite{obrazt,cdf}: ${\rm \kappa^{2}_{\gamma}<0.176}$ 
and ${\rm \kappa^{2}_{\rm Z}<0.533}$.

In principle, a large FCNC coupling could not only lead to the
associated production of a top plus a light quark at LEP2, but also to
sizable branching ratios of the top quark into $\mrm{\gamma c(u)}$ or
$\mrm{Zc(u)}$. This analysis uses only the ${\rm t \rightarrow bW}$
channel.
The reduction of the branching ratio BR(${\rm t \rightarrow bW}$) due
to possible FCNC decays is taken into account in the results section.

\section{Data and Monte Carlo Samples}
\label{sect:detector}

%=====================================================================%

The present analysis is based on data collected by the OPAL
detector~\cite{detector} from 1998 to 2000 at centre-of-mass energies
between 189\,GeV and 209\,GeV. OPAL
is a multipurpose high energy physics detector incorporating excellent
charged and neutral particle detection and measurement capabilities.
The search presented here uses 600.1\,\pb\ of data collected at high
energies for which the necessary detector components were required to
be operational while the data were recorded.  In addition, $11.3\,\pb$
of calibration data were collected at $\sqrt{s} \sim \mZ$ in
1998--2000 and have been used for fine tuning of the Monte Carlo
simulation. 
In this
paper, the data sample recorded in 1998 at $\sqrt{s} \simeq 189$\,GeV
is analysed in one bin, while the data from 1999 are divided into four
samples at $\sqrt{s} \simeq$ 192, 196, 200 and 202\,GeV. The data collected
in 2000 is analysed in two samples of mean centre-of-mass energies of
about 205 and 207\,GeV.

A variety of Monte Carlo samples were generated for the evaluation
of the detection efficiencies for single top production and SM
background processes.  In all samples, the hadronisation process is
simulated with {\tt JETSET} 7.4~\cite{pythia} with parameters described in
Reference~\cite{opaltune} and the \Wboson\ boson mass is set to $\mW =
80.33$\,GeV$/c^{2}$.  For each Monte Carlo sample, the detector
response to the generated particles is simulated by a
{\tt GEANT3} based package~\cite{gopal}.

The main Monte Carlo generator used for the description of our signal is
{\tt PYTHIA}~\cite{pythia}, which produces ${\rm \bar{t}c(u)}$ via an
s-channel exchange of a $\mathrm{Z}$ boson. The top quark decays into
a b quark and a W boson before it can form a bound state or radiate
gluons.  A colour string is formed between the ${\rm \bar{b}}$ 
and ${\rm c(u)}$ quarks to
form a colour singlet. 
All couplings and quark fragmentation parameters for ${\rm
e^+e^-\rightarrow\bar{t}c(u)}$ are set as in Z decays to quark pairs.
For an evaluation of systematic errors associated with the Monte Carlo
modelling, the signal is also modelled with a different {\tt PYTHIA}
process and the {\tt EXOTIC} generator~\cite{exotic}. This other {\tt
PYTHIA} process is based on a model~\cite{BEN85a} for the production
of a horizontal gauge boson, called ${\rm R^0}$, with the decay ${\rm
R^0} \to {\rm \bar{t}c(u)}$. The {\tt EXOTIC} generator was developed
for pair or single production of heavy and excited fermions. Here, the
top quark is the heavy fermion and its production is associated with a
c or u quark.  A sequential decay model is assumed with all couplings
to the known gauge bosons set to the SM expectations. For all three
generators, samples for three different top quark masses (169, 174,
and 179\,GeV$/c^{2}$) have been generated.
The signal Monte Carlo samples used for the reaction ${\rm
e^+e^-\rightarrow\bar{t}c(u)}$ encompass a wide range of schemes for 
the form of the FCNC couplings, the angular distributions of the 
final state particles, and the parton shower parameters.

The background processes are simulated, with more statistics than
the data collected, using the following event generators: {\tt
PYTHIA}, {\tt KK2F}~\cite{kk2f}, and {\tt HERWIG}~\cite{herwig} for
\Zgs \ra\ \qq($\gamma$); grc4f~\cite{grc4f}, {\tt
KORALW}~\cite{bib:KORALW}, and {\tt EXCALIBUR}~\cite{excalibur} for
four-fermion (4f) processes; and {\tt HERWIG}, {\tt
PHOJET}~\cite{bib:phojet}, and {\tt Vermaseren}~\cite{bib:verma} for
two-photon scattering.

\section{Event Selection}
\label{sect:event}

The searches for single top events $\mrm{e^+e^-\rightarrow \bar{t}c(u)
\rightarrow \bar{b}Wc(u)}$ are sensitive to the leptonic and hadronic
decays of the \Wboson\ boson: $\Wlnu$ and $\Wqq$. The leptonic channel is a
clean final state with specific topology and kinematics; it 
is characterised by two hadronic jets, one isolated lepton,
some missing energy (carried away by the neutrino), and the presence
of a b-hadron decay. While the hadronic channel is not as clean as
the leptonic channel, it is statistically significant because BR${\rm
(W \rightarrow hadron)\approx}$ 68\% and BR${\rm (W \rightarrow
\ell\bar{\nu_{\ell}})\approx}$ 32\% ($\ell=e$, $\mu$, 
and $\tau$). The hadronic channel is
characterised by four hadronic jets with specific topology and
kinematics, large visible energy, and the presence of a b-hadron
decay. 
Common search procedures are applied to both channels. The event
selection begins with loose global preselection criteria designed to
remove most of the two-photon and low multiplicity events.  To obtain
optimal resolution for single top candidates, kinematic fits are
performed to reject badly reconstructed events and background which
are not compatible with the topology of single top events.
Consequently, the event selection is followed by detailed preselection
cuts for both the leptonic and the hadronic channels. The final
candidate events are then identified using relative likelihood
functions.  Each step will be described briefly in the following 
subsections.

%-------------------------------------------------------------------
\subsection{Global Event Selection Criteria}
\label{sect:global}

Events are reconstructed from tracks in the central
tracking system and energy clusters in the electromagnetic and hadron
calorimeters, using selection criteria which are the same as
those used for the OPAL Higgs analysis~\cite{higgsold}. Because of the
presence of jets in a single top event, general multi-hadronic
preselections are applied. Each event must qualify as a
multi-hadronic final state according to the criteria of
References~\cite{tkmh,l2mh}.
These cuts remove events with low multiplicity or little visible
energy and  reject effectively  two-photon and pure leptonic
events.

The final state particles and clusters are grouped into jets using the Durham
algorithm~\cite{durham}. These jets are used as reference jets in the
following assignment procedure.  In calculating the visible energies
and momenta, $E_{\mathrm vis}$ and $\vec{p}_{\mathrm vis}$, of the
event and of individual jets, corrections are applied to prevent
double-counting of the energy of the tracks and their associated
clusters~\cite{mt300}.

\subsection{Lepton Identification}
\label{sect:leptonid}

Lepton identification for the leptonic channel relies primarily on the
isolation criteria of a prompt charged particle.  Isolated leptons are
identified using the Neural Network (${\rm NN_{\ell}}$) described in
Reference~\cite{H:pn183}. The ${\rm NN_{\ell}}$ uses all tracks in an
event with $|\vec{p}|>2$\, GeV/$c$ which are considered one-by-one in
decreasing order of momentum. They are used as seed tracks and
all tracks and unassociated clusters within $10^{\circ}$ of the
seed track define the lepton.
Afterward the leptons are classified as one-prong or three-prong
candidates depending on the number of tracks within the $10^{\circ}$ 
cone.
Around the seed candidate an annular cone of $30^{\circ}$ is drawn
concentric with and excluding the $10^{\circ}$ narrow cone. This serves 
to define the isolation criteria of the lepton candidate. The ${\rm NN_{\ell}}$
provides a distinctive signature for high energy leptons from the particle
flow in the annular and narrow cones. However, this procedure is flavour
blind; the main interest is to retain high identification efficiency.
Thus, the ${\rm NN_{\ell}}$ topological identification is sensitive to the
detection of electrons, muons, and taus with efficiencies of 84\%,
84\%, and 75\%, respectively.  The probability of misidentifying a
hadron from a parton shower as a lepton is around 1\% for ${\rm
NN_{\ell}>0.75}$. The main source of misidentified leptons comes from
low-multiplicity gluon jets.

The lepton with the largest ${\rm NN_{\ell}}$ output in every event is
taken to be the lepton of the ${\rm \bar{t} \to \bar{b} W \to
\bar{b}\,\lnu}$ decay. In order to improve the performance of the
kinematic fits, a simple identification is used to determine the mass
(flavour) of the lepton candidate.  First, all three-prong candidates
are classified as taus.  Then, a lepton is classified as an electron
if ${\cal P}_{e}\geq0.5$, $E_{\ell}>20$\,GeV, and
$\cos\theta_{\ell-\nu}<0.25$, where ${\cal P}_{e}$ is the standard
OPAL electron identification probability~\cite{bib:NNlep}, $E_{\ell}$
is the energy of the lepton, and $\theta_{\ell-\nu}$ is the
opening angle between the lepton and the missing momentum vector. Of
the remaining leptons, the candidates with ${\cal P}_{e}<0.5$,
$E_{\ell}>20$\,GeV, and $\cos\theta_{\ell-\nu}<0.25$ are classified as
muons, while the others are labelled as taus.

%-------------------------------------------------------------------
\subsection{Event Kinematics}
\label{sect:kin}

At a centre-of-mass energy of $\sqrt{s} \approx 189$ GeV the top quark
is produced close to threshold. As the
top quark is nearly at rest, the ${\rm W}$ boson in the
$\mrm{e^+e^-\rightarrow \bar{t}c(u) \rightarrow \bar{b}Wc(u)}$
reaction has almost constant energy 
$E_{\rm W} \simeq (m_{\rm t}^2+m_{\rm W}^2-m_{\rm b}^2) / 2\,m_{\rm t}$, 
which also leads to fixed energy for the b quark 
$E_{\rm b} \simeq (m_{\rm t}^2-m_{\rm W}^2+m_{\rm b}^2) / 2\,m_{\rm t}$. 
With increasing centre-of-mass energy this unique kinematic signature gets 
diluted.

These specific kinematic properties are exploited by using kinematic
fits.  First, the event is constrained to pass a 4C kinematic fit
which ensures that the energy and momentum are conserved\footnote{For
semileptonic events, there are three unmeasured variables corresponding to
the neutrino momentum so that the effective number of constraints
in the leptonic mode is one. Nevertheless, any fit which implies that 
energy and momentum are conserved for both the leptonic and hadronic 
channels will be referred to as a 4C fit.}.  The 4C kinematic fit is
employed to remove badly reconstructed events and events with missing
particles along the beam pipe.  The $\chi^2$ probability of the 4C fit
is thus required to be larger than $\mrm{10^{-5}}$.

To obtain optimal resolution for the reconstructed candidates and
performance for the jet assignment, we use additional kinematic
fits which enforce energy and momentum conservation and impose the
appropriate mass constraints.  These fits are referred to as the 6C
fits with the ${\rm \bar{t}c(u)}$ or the \WW\ hypothesis:

\begin{itemize}
\item {\bf\boldmath{${\rm e^+e^- \to \bar{t}c(u) \to \bar{b} W
c(u)}$:}} \Wboson\ boson and top quark invariant mass constraints.

 \item {\bf\boldmath{${\rm \eeWW}$:}} two W boson invariant mass constraints.

\end{itemize}

In the 6C kinematic fits the \Wboson\ mass and the top quark mass are
fitted with a soft constraint, approximating the Breit-Wigner shapes
by Gaussian resolution functions.
As for the 4C fit, we still refer any mass constrained fit as a 6C 
fit for ${\rm \bar{b}\,\lnu\,c(u)}$ events.
To ensure that the kinematic properties of the event candidates match
our signal process, we require ${\cal P}({\mbox{6C}})>10^{-5}$ for 
the ${\rm \bar{t}c(u)}$ 6C fit.

%-------------------------------------------------------------------
\subsection{B-Tagging}
\label{sect:btag}

The dominant background in this analysis comes from \WW\ events. In
\eeWW\ events, the only heavy quark commonly produced is the charm
quark. The production of bottom quarks is  highly suppressed
due to the small magnitude of \Vub\ and \Vcb\ and the large mass of
the top quark.  Furthermore, since the top quark is expected to decay
into a $\mrm{bW}$ pair, the tagging of jets originating from b quarks
plays an important role in single top production searches. The
jet-wise b-tagging algorithm, which has been developed for the Higgs
boson search, uses three independent b-tagging methods: (1) lifetime
tag, (2) high-$p_{\mathrm{T}}$ lepton tag, and (3) jet shape tag.
These three methods are combined using an unbinned likelihood method
to form a single discriminating variable for each jet~\cite{H:pn183}.
The b-tag becomes important for higher centre-of-mass energies because
the kinematic situation changes and the signal is less
well separated from the \WW\ background.

%-------------------------------------------------------------------
\subsection{Jet Assignment}
\label{sect:jet}

In the hadronic channel, the correct assignment of particles to jets
plays an essential role in reducing four-jet like backgrounds. There
are twelve possible combinations to assign two jets to the \Wboson\
boson, the third jet to the ${\rm \bar{b}}$ quark, and the fourth jet
to the light flavoured quark.  Therefore a discriminating variable is calculated,
which is a combination of the 6C kinematic fit probability and the
b-tag variable, in order to find the best matching combination to
the signal hypothesis.  The 6C fit helps to identify the two jets
coming from the \Wboson\ and to find the third jet which matches
kinematically to form the invariant top quark mass. In addition the
b-tag variable helps to identify if this latter jet is a b-jet.
The jet assignment which yields the largest
${\cal P}({\mbox{b-tag,\,6C}})$ is used to choose the jet/quark
assignment.  ${\cal P}({\mbox{b-tag,\,6C}})$ is calculated as:
$$
{\cal P}({\mbox{b-tag,\,6C}}) = \frac{{\cal P}({\mbox{6C}})\,{\cal P}
({\mbox{b-tag}})}
{{\cal P}({\mbox{6C}})\,{\cal P}({\mbox{b-tag}}) + 
[1-{\cal P}({\mbox{6C}})]\,[1-{\cal P}({\mbox{b-tag}})]},
$$
where ${\cal P}({\mbox{b-tag}})$ is the b-tag variable and ${\cal
P}({\mbox{6C}})$ is the probability from the ${\rm \bar{t}c(u)}$ 6C
fit. 

In the leptonic channel, the correct jet/quark assignment plays an
important role in reducing signal-like background topology.  In ${\rm e^+e^-
\to \bar{t} c(u) \to \bar{b}\,\lnu\,c(u)}$, there are only two
possible jet assignments. One of the jets must come from the
hadronisation of the ${\rm \bar{b}}$ quark and the other from the
light flavoured quark.  The bottom jet is taken to be the one with the
largest ${\cal P}({\mbox{b-tag,\,6C}})$ from the
${\rm \bar{t}c(u)}$ 6C fit.

With the jet assignment method described here, a Monte Carlo study
shows that the rate of correct b-jet (non b-jet) assignment at \roots\
= 189\,GeV is about 96\% (94\%) and 84\% (73\%) for the leptonic and
hadronic channels, respectively.

%-------------------------------------------------------------------

%%%%%%%%%%%%%%%%%%%%%%%%%%%%%%% PRESELECTION %%%%%%%%%%%%%%%%%%%%%%%%%%
\subsection{Single Top Candidate Preselection}
\label{sect:presel}

To help further reduce the background after the global event selections,
the kinematic fits, the b-tagging, and the jet assignment, individual
preselection criteria are enforced for both the leptonic and the 
hadronic channels.

\subsubsection{Preselection: Leptonic Channel}
\label{sect:prelep}

The following preselection cuts are applied in order to reduce
background with a different topology to our signal process:

\begin{enumerate}

 \item $N_{\rm lepton} \geq 1$, where $N_{\rm lepton}$ is the number 
 of lepton candidates as described in Section~\ref{sect:leptonid}.
 
 \item $|\cos\theta_{\rm miss}|<0.9$, where $|\cos\theta_{\rm miss}|$
 is the cosine of the polar angle of the missing momentum vector. This
 cut rejects a large portion of the \qq$(\gamma)$ background.

 \item $M_{\rm vis}/\roots > 0.20$, where $M_{\rm vis}$ is the invariant 
 mass calculated from the visible energy $E_{\rm vis}$ and the visible 
 momentum $\vec{p}_{\mathrm vis}$ of the event. 

 \item $|\vec{p}_{\mathrm miss}|/\roots < 0.50$, where
 $\vec{p}_{\mathrm miss}$ is calculated from the visible momentum
 $(\vec{p}_{\mathrm miss}=-\vec{p}_{\mathrm vis})$. This cut reject
 events with large missing momentum, such as \qq$(\gamma)$ background
 when the photon escapes detection.

 \item $0.20<\sum{|\vec{p}_{\rm T}|}/\roots < 0.90$, where $\sum{|\vec{p}_{\rm T}|}$ 
 is the scalar
 sum of the transverse momentum components for all the good tracks and 
 unassociated clusters. This cut prevents the visible momentum
 being toward the beam direction and rejects non-radiative 
 \qq\ events with no missing energy.

 \item ${\rm NN_{\ell}}>0.75$, where ${\rm NN_{\ell}}$
 is the primary lepton Neural Network output as described in 
 Section~\ref{sect:leptonid}.
 
\end{enumerate}

After the preselection the background is well described by 4f and \qq\
events. Other final states, such as two-photon events, are
negligible. The main background (around 95\%) is due to \WWqqln\ events. The
fraction of events with four quarks in the final state selected with
the leptonic preselection criteria is negligible.

%==========================================================================%
\subsubsection{Preselection: Hadronic Channel}
\label{sect:prehad}

The following  preselection cuts are applied in order to select only
four-jet like events:

\begin{enumerate}

\item The event must contain at least 15 charged tracks.

\item The maximum energy of any electron or muon found in the 
event (identified as described in Reference~\cite{opal:lepton}) 
must be less than 40\,GeV.

\item The radiative process \epm $\rightarrow \Zgs \gamma {\rm
\rightarrow q\bar{q}\gamma}$ is reduced by requiring that the
effective centre-of-mass energy $\sqrt{s^{\,\prime}}$~\cite{l2mh} be at
least 150\,GeV.

\item The Durham jet resolution parameter $y_{34}$, at which the
number of  jets changes from three to four, is required to be larger than
0.001.

\item The \Zgs ${\rm \rightarrow q\bar{q}}$ background is further
suppressed by requiring that the event shape parameter
$C$~\cite{cpar}, which is close to one for spherical events, is larger 
than 0.4.

\end{enumerate}

After the preselection the background is well described by 4f and \qq\
events.  The expected background is composed of 41\% (70\%) of 4f and
59\% (30\%) of \qq\ processes at $\sqrt{s} \simeq$ 189 (207)\,GeV.
Other final states, such as two-photon events, are
negligible. The fraction of \qqln\ events selected by the hadronic
preselection criteria is less than 1\%.

%%%%%%%%%%%%%%%%%%%%%%%%%%%%%%% FCNC likelihood %%%%%%%%%%%%%%%%%%%%%%%%%%

\subsection{Likelihood Selection}
\label{sect:likelihood}

The final separation of the signal from the background is achieved
with a conventional multi-variable relative likelihood function~\cite{karlen}:
$$
{\cal L} = \frac{ {\cal P}_{\rm signal} } 
{ {\cal P}_{\rm signal} + {\cal P}_{\rm background} }, 
{\mbox{~~with~~}} {\cal P} = \prod_{i} p_{i}.
$$
The template (or reference) histograms of the input variables, $p_i$, are used
as the probability density functions for the calculation of ${\cal P}_{\rm signal}$ 
and ${\cal P}_{\rm background}$. We rely on Monte Carlo events to compute the 
probability density functions.

\subsubsection{Likelihood: Leptonic Channel}
\label{sect:liklep}

For each event satisfying the \qqln\ preselection cuts, a binned
likelihood function is constructed, with one class for the signal and
one for the 4f background. The relative likelihood is calculated using
the following variables:
\begin{description}

  \item  [\boldmath $E_{c(u)}$:] The energy of the light flavoured jet.

  \item  [\boldmath $M_{qq}^{\rm 4C}$:] The invariant mass of the 
  \qq\ system after the 4C fit.

  \item  [\boldmath 
  $M_{\ell\nu} = \sqrt{ E_{\rm beam}^2 - (\vec{p}_{\ell\nu}^{\rm \,4C})^2}$:]  
  Pseudo mass of the $\lnu$ system after the 4C fit, calculated from the 
  beam energy and the momentum of the $\lnu$ pair. 

  \item  [\boldmath $\ln y_{12}$:] The logarithm of the Durham jet resolution 
  parameter at which the number of reconstructed jets passes from one to two.

  \item  [\boldmath $M_{qq}^{\rm 6C} + M_{\ell\nu}^{\rm 6C}$:] 
  The sum of the di-jet and $\lnu$ invariant masses for the 6C fit under 
  the \myWW\ hypothesis.

  \item  [{\bf b-tag:}] The b-tag variable of the selected bottom jet.

  \item  [{\bf\boldmath ${\cal P}$(b-tag,\,6C):}] The discriminant variable
  which combines the b-tag variable and the ${\rm \bar{t}c(u)}$ 6C fit probability.

\end{description}

\begin{figure}[p]
\centerline{\epsfig{file=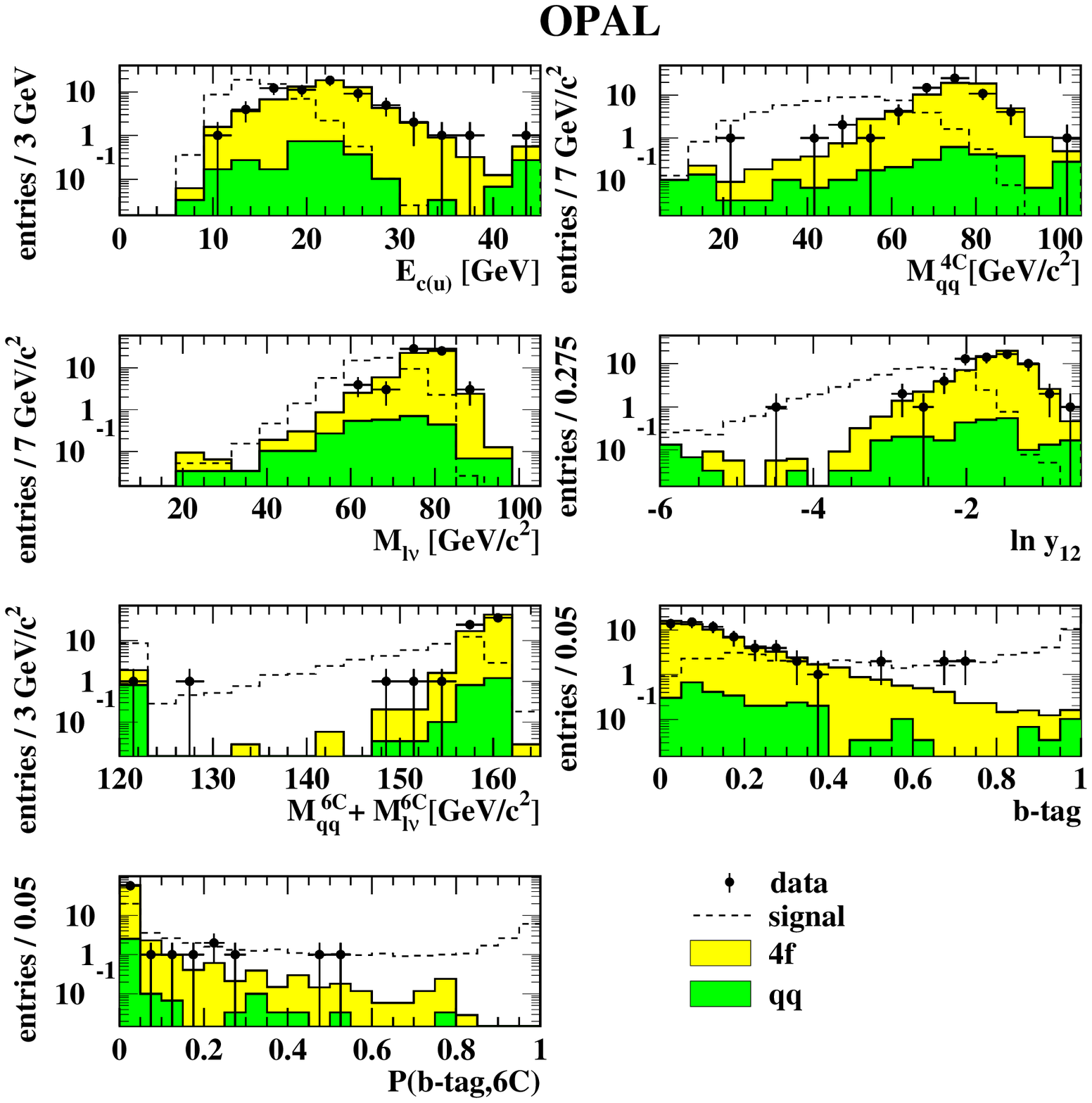,width=1.0\textwidth}}
\caption[]{\label{plot:linputlept}Distributions of the likelihood
        variables for the leptonic channel at $\sqrt{s} \simeq 189$\,GeV.  
        Comparisons between the data,
        the SM 4-fermion (light grey), and \qq\ backgrounds (grey) are
        shown.  The dashed line represents single top MC events with \mtop\ =
        174\,$\mrm{GeV/c^{2}}$ and an arbitrary cross-section of
        $\sigma_{\mathrm{top}}$ = 3\,pb.}
\end{figure}

Jets tagged as light flavoured jets in the background from SM
processes have much higher values of $E_{c(u)}$. The second and third
variables offer good discrimination for \WWqqln\ events since they
exploit the specific angular distribution of signal events. The variable
$y_{12}$ exploits the unique jet distribution in ${\rm \bar{t}c(u)}$
events. To further remove the background from semileptonic \myWW\
decays, we use $M_{qq}^{\rm 6C} + M_{\ell\nu}^{\rm 6C}$. Finally, we
use the b-tag variable and the ${\cal P}$(b-tag,\,6C) of the b-jet to
separate bottom-less events.  Figure~\ref{plot:linputlept} shows the
distributions of the input variables for data, the SM background, and
the simulated single top signal at $\sqrt{s} =189$\,GeV.  There is
good agreement between data and Monte Carlo distributions from
background processes.

\subsubsection{Likelihood: Hadronic Channel}
\label{sect:likhad}

For each event satisfying the \qqqq\ preselection cuts, a binned
likelihood function is constructed, with one class for the signal and
two for the \qqb \qqb\ and the \qq\ backgrounds. The relative
likelihood is calculated using the following variables:

\begin{description}

  \item [\boldmath $\chi^2(6\rm{C\ fit})$:] 
    The $\chi^2$ of the ${\rm \bar{t}c(u)}$ 6C kinematic fit.

  \item [\boldmath $E_{\rm c(u)}/E_{\rm vis}$:]
   The ratio of the energy of the $\mrm{c(u)}$ jet and the total visible 
   energy.

  \item [Thrust:] The value of the thrust for the event~\cite{pythia}.

  \item  [{\bf b-tag:}] The b-tag variable of the selected bottom jet.

  \item [\boldmath $\cos(\angle(\vec{p}_{\rm Wq1},\vec{p}_{\rm Wq2}))$:]
    The cosine of the angle between the two jets tagged
    as decay products of the ${\rm W}$ boson.

\end{description}

\begin{figure}[p]
\centerline{\epsfig{file=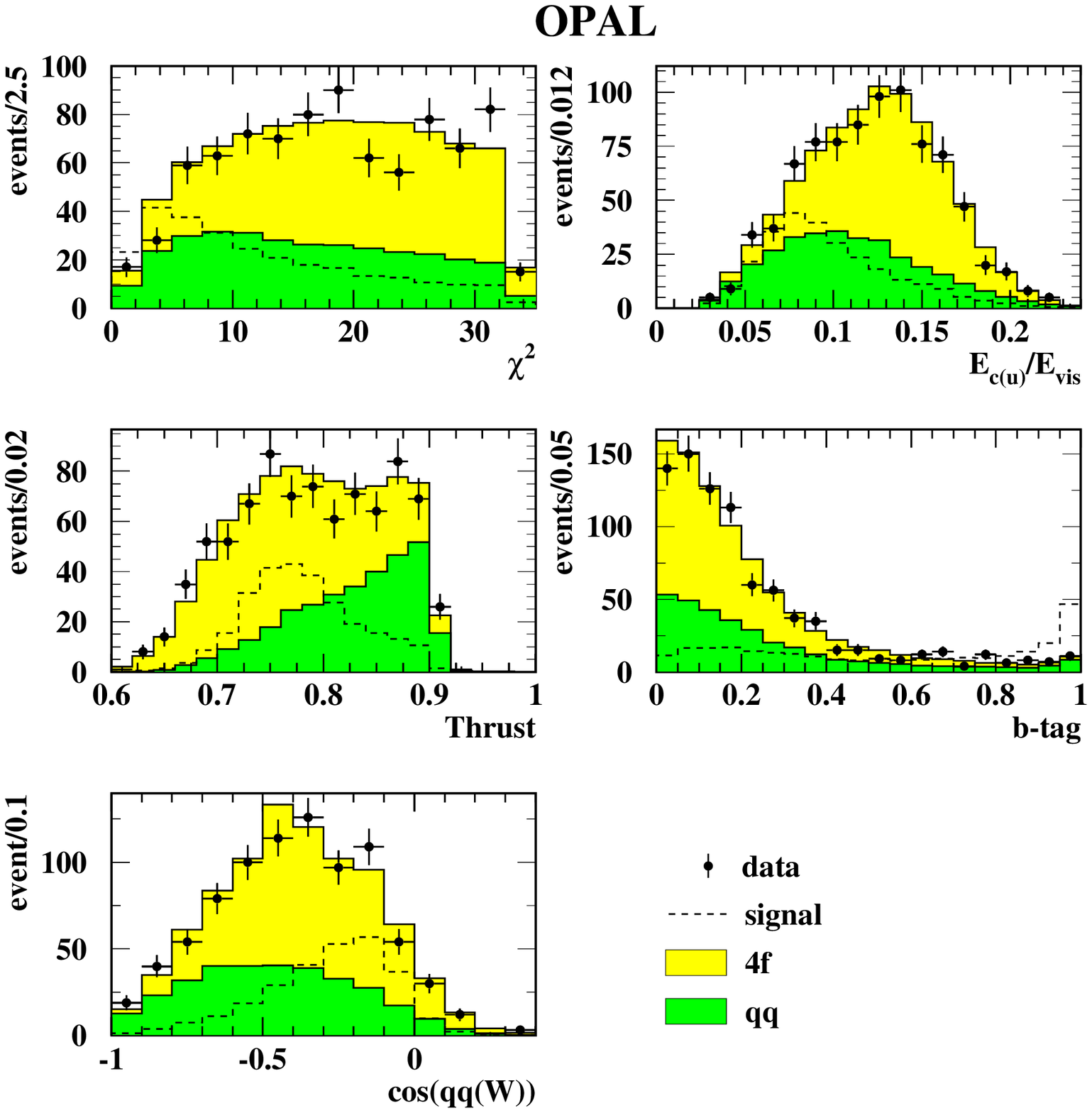,width=1.0\textwidth}}
\caption[]{\label{plot:linputhad}Distributions of the likelihood
        variables for the hadronic channel at $\sqrt{s} \simeq 189$\,GeV.  
        Comparisons between the data,
        the SM 4-fermion (light grey), and \qq\ backgrounds (grey) are
        shown.  The dashed line represents single top MC events with \mtop\ =
        174\,$\mrm{GeV/c^{2}}$ and an arbitrary cross-section of
        $\sigma_{\mathrm{top}}$ = 3\,pb.}
\end{figure}

The thrust variable exploits the different event topologies between
signal and backgrounds. The b-tag variable is used as an effective
likelihood input because the top quark is expected to decay into a
{\rm b} quark.  The other three variables exploit the specific
kinematics of signal events.  Figure~\ref{plot:linputhad} shows the
distributions of the input variables for data, the SM background, and
the simulated single top signal at $\sqrt{s} =189$\,GeV.  There is
good agreement between data and Monte Carlo distributions from
background processes.

\subsubsection{Likelihood: Both Channels}
\label{sect:likboth}

\begin{figure}[p]
\centerline{\epsfig{file=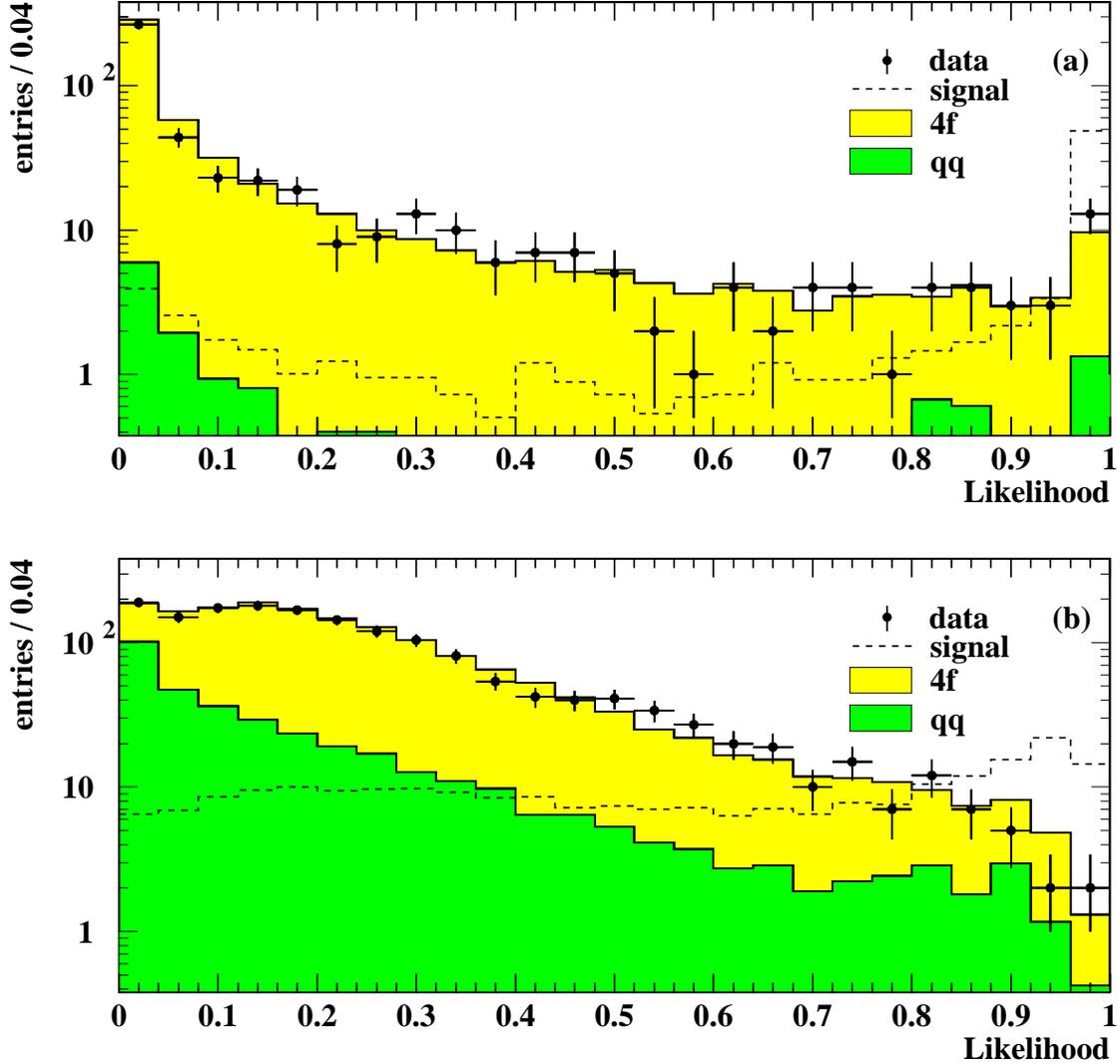,width=1.0\textwidth}}
\caption[]{\label{plot:likeout}Distributions of the likelihood
        variables for (a) the leptonic and (b) the hadronic channels. The
        comparison between the data collected in 2000 at 
        $\sqrt{s} \simeq 205$\,GeV and $\sqrt{s} \simeq 207$\,GeV, the SM 
        4-fermion (light grey), and the
        $\mathrm{q}\bar{\mathrm{q}}$ backgrounds (grey) is shown.  The dashed
        line represents single top MC events with $m_{\mathrm{t}}$ =
        174\,$\mathrm{GeV/c^{2}}$ and an arbitrary cross-section of
        $\sigma_{\mathrm{top}}$ = 3\,pb.} 
\end{figure}

In Figure~\ref{plot:likeout} the relative likelihood functions for the
leptonic and hadronic channels are shown for data collected at
$\sqrt{s} \simeq$ 205 - 207\,GeV and for the SM expectation. No excess
of events is observed. The likelihood functions for FCNC signal events
for an assumed arbitrary cross-section of 3\,pb are also depicted in
Figure~\ref{plot:likeout}. It can be seen that the expected signal
contribution is concentrated at high values of ${\cal L}$. 

There is no evidence of single top quark production in the data for 
any \roots. Thus, the final likelihood cuts are chosen at each value 
of \roots\ so as to minimise the expected
upper limit on the signal cross-section and thus to maximise
the expected exclusion sensitivity. The number of selected data 
and expected SM background events as a function of the 
centre-of-mass energies are shown in Table~\ref{tab:result}.

%--------------------------------------------------------------------------
%%%%%%%%%%%%%%%%%%%%%%%%%%
\begin{table}[ht]
\begin{center}
\begin{tabular}{|c|c|c|c|c|c|c|}
\hline
Label & $\sqrt{s}$ & Lumi. & 
\multicolumn{2}{|c|} {Leptonic Channel} &
\multicolumn{2}{|c|} {Hadronic Channel}           \\
(GeV) & (GeV) & 
(pb$^{-1}$) & 
Data & SM Total  &
Data & SM Total  \\
\hline
\hline
189 & 188.7 & 172.1  &  3  &    4.0   &   13  &  11.6 \\
192 & 191.6 &  28.9  &  0  &    1.0   &    7  &   5.1 \\
196 & 195.6 &  74.8  &  1  &    2.9   &    6  &   6.4 \\
200 & 199.6 &  77.2  &  3  &    2.7   &   10  &   9.4 \\
202 & 201.6 &  36.1  &  2  &    1.2   &    8  &   7.5 \\
205 & 205.1 &  80.3  &  1  &    2.0   &   11  &  10.1 \\
207 & 206.8 & 130.8  &  6  &    3.8   &   14  &  16.4 \\
\hline
\end{tabular}
\caption{The luminosity-weighted mean centre-of-mass energies,
the integrated luminosities, the number of selected data and 
expected SM background events
at $\sqrt{s}$ = 189\,-\,209\,GeV are shown
for the leptonic and hadronic channels.}
\label{tab:result}
\end{center}
\end{table}
%%%%%%%%%%%%%%%%%%%%%%%%%%

%%%%%%%%%%%%%%%%%%%%%%%%%%%% SYSTEMATIC ERRORS %%%%%%%%%%%%%%%%%%%%%%%%%%
\section{Systematic Errors}

\subsection{Signal Efficiencies and SM Backgrounds}
\label{sec:syst}

Sources of systematic uncertainties are investigated for their effect
on the signal detection efficiencies and the SM backgrounds.  They are
listed in Table~\ref{tab:syserr} for three of the energy bins
and are discussed below.  
All checks were performed for all centre-of-mass energies.
Possible color reconnection and Bose-Einstein effects were not
investigated.

The errors on the background and signal rates from the modelling of
the preselection variables and of the detector response are a few
percent.
These uncertainties are evaluated based on comparisons of the
distributions of the variables in the calibration data collected at
$\sqrt{s} \sim \mZ$ and the Monte Carlo simulation.  The
effects of detector mis-calibration and deficiencies were investigated
by varying the jet and lepton energy scales over a reasonable
range~\cite{bib:Wmass}. The uncertainties on the energy resolution and
the angular resolution were also evaluated, but have much smaller
effects. A comparison of alternative Monte Carlo generators for the
background accounts for an additional
uncertainty on the background rates. The difference between the
luminosity-weighted centre-of-mass energies in data and the value of
$\sqrt{s}$ used in the main Monte Carlo samples results in an
additional uncertainty on the background and signal selection
efficiencies due to the use of an energy constraint in the kinematic fits.
Lepton identification accounts for an extra uncertainty for the
leptonic channel.

One of the dominant errors in both analysis channels arises from the
b-tagging.  Recent improvements in the knowledge of heavy quark
production processes and decays, such as the b-hadron charged decay
multiplicity and the gluon splitting rate to heavy quarks, are taken
into account in the analysis by reweighting Monte Carlo
events~\cite{opal:higgs00}.  The sensitivity to the b-vertex
reconstruction was assessed by degrading or improving the tracking
resolution in the Monte Carlo.  It was found that changing the track
parameter resolutions by $\pm5\%$ in the Monte Carlo simulation covers
the range of possible differences between data and simulated events.
Overall it leads to an uncertainty of 3.8-8.4\% for the b-tag rates of
background and signal events.  The finite size of the Monte Carlo
samples used in this analysis results in an additional uncertainty of
a few percent for the background and the signal selection
efficiencies.

All the different systematic effects for the background and the signal
efficiencies are treated as being independent. The total uncertainties
on the background and signal rates, for both  the leptonic and the
hadronic channel, are in the same range and show small dependencies on
the centre-of-mass energy. For each centre-of-mass energy, the systematic
errors are included in the calculation of the cross-section upper limits.

%-------------------------- TABLE SYSTEMATICS --------------------------
\begin{table}[t]
\begin{center}
\begin{tabular}{|c||c|c||c|c|}
\hline
Source of                                &  
\multicolumn{2}{|c||} {Leptonic Channel}  &
\multicolumn{2}{|c|} {Hadronic Channel} \\
Systematic Error     
& $\Delta$efficiency & $\Delta$background & $\Delta$efficiency & $\Delta$background \\
\hline\hline
Preselection         &  1.0/1.4/1.2  & 2.0/2.2/1.9  & 1.0/0.4/0.3  &  3.2/1.4/0.5  \\
Detector Response    &  1.0/2.2/3.4  & 1.4/1.7/1.2  & 0.6/2.0/1.5  &  1.0/1.0/3.0  \\
Background           &   - / - / -   & 6.8/7.4/6.9  &  - / - / -   &  5.0/5.0/5.0  \\
\roots\ in MC        &  1.1/2.4/1.9  & 2.7/2.1/2.3  & 0.6/0.8/0.5  &  1.5/1.3/1.4  \\
Lepton ID            &  4.4/5.0/4.8  & 3.0/4.0/3.5  &  - / - / -   &  - / - / -    \\
b-tagging            &  4.2/6.6/5.2  & 7.8/5.6/7.0  & 3.8/5.3/5.2  &  6.9/5.5/8.4  \\
MC Statistic         &  2.2/2.3/2.1  & 1.5/2.0/1.6  & 2.0/1.8/1.8  &  5.4/5.0/4.8  \\
FCNC modelling       &  7.2/8.3/3.5  &  - / - / -   & 7.9/6.6/5.0  &  - / - / -    \\ 
\hline
Total               &  9.8/12.5/9.1  & 11.5/10.9/11.0 & 9.1/8.9/7.6 & 10.7/9.2/11.4 \\ 
\hline
\end{tabular}
\end{center}
\caption[]{\label{tab:syserr}
  The relative systematic errors (in \%) on the signal reconstruction
  efficiency and on the background modelling for 
  \roots\ = 189/200/207\,GeV.}
\end{table} 
%-------------------------------------------------------------------

\subsection{FCNC Modelling}

Several methods of producing FCNC can be compared.  A comparison of
the results obtained with the {\tt EXOTIC} and the {\tt PYTHIA}
samples described in Section~\ref{sect:detector} allows an estimate of
the uncertainty due to the model used for the signal process.  The
difference is taken as a modelling uncertainty on the simulation of
signal events.  It is summarised in Table~\ref{tab:syserr}.  The main
disparities between all the generator schemes are the angular
distributions of the particles produced in the final state and the
parton shower modelling of the initial quarks. This latter effect
gives rise to one of the largest uncertainties on the signal
reconstruction efficiency.

\subsection{Top Quark Mass}
The largest systematic uncertainty comes from the sensitivity
of the event selection to the assumed value of the top quark mass.
In the analysis we assume the mass of the top
quark to be 174\,GeV$/c^{2}$. To take this dependency into 
account, the variation of the reconstruction
efficiency is investigated using Monte Carlo events with $m_{\rm t}$=
169 and 179\,GeV$/c^{2}$ incorporating the experimental systematic and
the FCNC model uncertainties described in the previous sections.  The
dependence of the reconstruction efficiencies on the top quark mass
for the leptonic and hadronic channels is summarised in
Table~\ref{tab:effxsec}.

%--------------------------------------------------------------------------
%%%%%%%%%%%%%%%%%%%%%%%%%%
\begin{table}[ht]
\begin{center}
\begin{tabular}{|c|c|c|c|c|c|c|c|c|c|}
\hline
$\sqrt{s}$ & 
\multicolumn{3}{|c|} {\mtop\ = 169\,GeV$/c^2$} &
\multicolumn{3}{|c|} {\mtop\ = 174\,GeV$/c^2$} &
\multicolumn{3}{|c|} {\mtop\ = 179\,GeV$/c^2$} \\
(GeV) & 
$\epsilon_{\ell}$ & $\epsilon_{q}$ &
%${\rm \sigma_{95}^{exp.}}$ & 
${\rm \sigma_{95}^{obs.}}$ &
$\epsilon_{\ell}$ & $\epsilon_{q}$ &
%${\rm \sigma_{95}^{exp.}}$ & 
${\rm \sigma_{95}^{obs.}}$ &
$\epsilon_{\ell}$ & $\epsilon_{q}$ &
%${\rm \sigma_{95}^{exp.}}$ & 
${\rm \sigma_{95}^{obs.}}$ \\
\hline
\hline
189              &     7.5    & 10.3      &  0.30 &   
                       9.1    & 12.8      &  0.24 &
                       6.1    & 10.0      &  0.33 \\
                                                      
192              &     7.5    & 15.3      &  0.99 &
                       9.5    & 18.0      &  0.81 &
                       6.9    & 14.9      &  1.04 \\
                                                      
196              &     7.1    & 12.8      &  0.39 &
                       8.7    & 14.7      &  0.33 &
                       7.2    & 12.1      &  0.40 \\
                                                      
200              &     7.1    & 14.7      &  0.55 &
                       8.0    & 16.0      &  0.50 &
                       7.0    & 15.1      &  0.55 \\
                                                      
202              &     6.6    & 17.7      &  1.00 &
                       7.5    & 18.6      &  0.93 &
                       6.9    & 17.3      &  1.00 \\
                                                      
205              &     5.9    & 14.4      &  0.48 &
                       7.0    & 15.7      &  0.43 &
                       6.2    & 13.9      &  0.49 \\
                                            
207              &     5.8    & 12.8      &  0.47 &     
                       6.7    & 15.4      &  0.40 &
                       6.1    & 13.6      &  0.45 \\

\hline
\end{tabular}
\caption{The reconstruction efficiencies for the leptonic
($\epsilon_{\ell}$) and the hadronic ($\epsilon_{q}$) channels are
shown. The overall
%expected (${\rm \sigma_{95}^{exp.}}$) and
measured 95\% \CL\ upper limits on single top production 
cross-section (${\rm \sigma_{95}^{obs.}}$) are reported. The statistical and
systematic uncertainties are included in the calculation of
the upper limits.  The efficiencies (in \%) and the limits
on the cross-section (in pb) are shown as a function of the
centre-of-mass energy for $m_{\rm t}=$ 169, 174, and 179\,GeV$/c^2$.
These results assume a 100\% branching fraction of the top quark into
b\,W.}
\label{tab:effxsec}
\end{center}
\end{table}
%%%%%%%%%%%%%%%%%%%%%%%%%%

%%%%%%%%%%%%%%%%%%%%%%%%%%%% RESULTS %%%%%%%%%%%%%%%%%%%%%%%%%%
\section{Results}
\label{sec:results}

No evidence for single top quark production is observed in \epm\
collisions at centre-of-mass energies between 189-209\,GeV.  Limits on
the single top cross-section have been derived at the 95\% \CL\ from
the measurements of the number of observed events, the reconstruction
efficiencies, and the integrated luminosities~\cite{junk}. The upper limit
calculations for each individual centre-of-mass energy are summarised
in Table~\ref{tab:effxsec}.  Those results include both the
statistical and systematic errors and are valid under the assumption
that \mtop\ = 169, 174, and 179\,GeV$/c^{2}$ and that BR(${\rm t \to
b\,W}$) = 100\%.  The CDF limits constrain the ${\rm t\rightarrow
\Vboson c(u)}$ FCNC branching ratio to be smaller than about 36\%~\cite{cdf}
for $\Vboson = \gamma$ or \Zboson, so that in a pessimistic scenario
the OPAL efficiencies and cross-section limits quoted in 
Table~\ref{tab:effxsec} should be rescaled by 64\%.  

The combination of all the data can be used to determine
limits on the anomalous coupling parameters $\kappa_{\gamma}$ and
$\kappa_{\rm Z}$. First, the QCD and the ISR effects
which modify the Born-level cross-section given in 
Equation~\ref{sintop:cross} must be considered.
The QCD correction is taken from Section 3 of Reference~\cite{qcdcor};
while the ISR correction is based on Reference~\cite{isrcor}.
Overall, the QCD and ISR corrections increase the 
Born-level cross-section by a constant factor of about 1.09
for all centre-of-mass energies and produce only
a small distortion to the OPAL exclusion region in the
$\kappa_{\gamma}-\kappa_{\rm Z}$ plane.

\begin{figure}[p]
\centerline{\epsfig{file=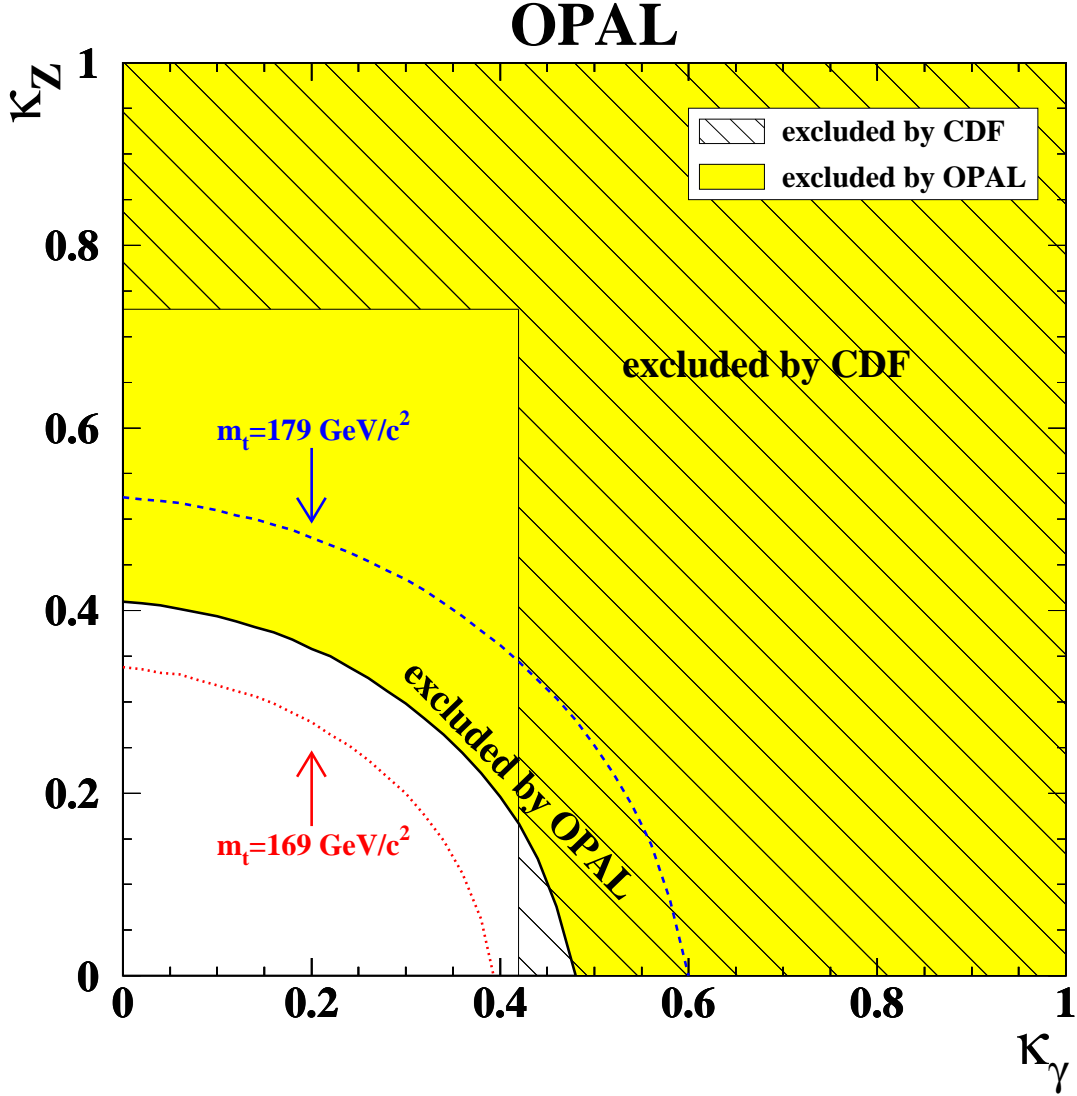,width=.95\textwidth}}
\caption[]{\label{plot:kgkz}The light grey region shows the OPAL
exclusion region at 95\% \CL\ in the $\kappa_{\rm Z}-\kappa_{\gamma}$
plane for $m_{\rm t}=$ 174\,GeV/$c^2$. The exclusion curves for
different values of top quark masses are also shown. The hatched area
shows the CDF exclusion region~\cite{cdf}. The OPAL limits include QCD and
ISR corrections to the Born-level cross-section defined in 
Equation~\ref{sintop:cross}.}
\end{figure}

The limits on the anomalous coupling parameters are obtained with the
likelihood ratio method described in Reference~\cite{junk}.  Each 
centre-of-mass energy for the leptonic
and the hadronic channel has been used as an independent
channel. The variation of the selection efficiencies for different 
top masses are taken from Table~\ref{tab:effxsec}.
Taking the statistical and systematic errors into account the
limit on the
anomalous coupling parameters in the $\kappa_{\gamma}-\kappa_{\rm Z}$
plane have been derived at the 95\% \CL.
The reduction of the branching ratio BR(${\rm t \rightarrow bW}$) due
to possible FCNC decays derived at each point in the
$\kappa_{\gamma}-\kappa_{\rm Z}$ plane is taken into account in this
generic FCNC production limit calculation. To compare our results with the
limits from CDF, exclusion regions for \mtop\ = 169, 174, and
179\,GeV$/c^{2}$ in the $\kappa_{\gamma}-\kappa_{\rm Z}$ plane were
obtained. The results are shown in Figure~\ref{plot:kgkz}.  
They correspond to upper limits of
$\kappa_{\gamma}<$0.48 and $\kappa_{{\rm Z}}<$0.41 for a top quark
mass of \mtop\ = 174\,GeV$/c^{2}$, which becomes
$\kappa_{\gamma}<$0.39 (0.60) and $\kappa_{{\rm Z}}<$0.34 (0.52) for
\mtop\ = 169 (179)\,GeV$/c^{2}$.  These exclusions translate into
branching fraction limits of $\mrm{Br(t \to \Zboson c}) + 
Br(\mrm{t \to \Zboson u})< 9.7/13.7/20.6$ \% for 
\mtop\ = $169/174/179$\,GeV$/c^{2}$. All these results are consistent with
recent results from the ALEPH Collaboration~\cite{topALEPH}.

\section{Summary}
A search for single top quark production via FCNC has been performed
with 600.1\,pb$^{-1}$ of data collected by OPAL in ${\rm e^+e^- }$
collision at $\sqrt{s}$ = 189\,-\,209\,GeV.  In total, 85 events were
selected in the data with a SM expectation of 84.1 events.  Limits on
single top quark cross-sections have been derived at the 95\%
\CL. This leads to model-dependent upper limits of
$\kappa_{\gamma}<$0.48 and $\kappa_{{\rm Z}}<$0.41 for a top quark
mass of \mtop\ = 174\,GeV$/c^{2}$.  The limits become
$\kappa_{\gamma}<$0.39 (0.60) and $\kappa_{{\rm Z}}<$0.34 (0.52) for
\mtop\ = 169 (179)\,GeV$/c^{2}$.  

%
%-----------------------------------------------------------------------
%
\appendix
\par
\section*{Acknowledgements:}
\par
We particularly wish to thank the SL Division for the efficient operation
of the LEP accelerator at all energies
 and for their close cooperation with
our experimental group.  We thank our colleagues from CEA, DAPNIA/SPP,
CE-Saclay for their efforts over the years on the time-of-flight and trigger
systems which we continue to use.  In addition to the support staff at our own
institutions we are pleased to acknowledge the  \\
Department of Energy, USA, \\
National Science Foundation, USA, \\
Particle Physics and Astronomy Research Council, UK, \\
Natural Sciences and Engineering Research Council, Canada, \\
Israel Science Foundation, administered by the Israel
Academy of Science and Humanities, \\
Minerva Gesellschaft, \\
Benoziyo Center for High Energy Physics,\\
Japanese Ministry of Education, Science and Culture (the
Monbusho) and a grant under the Monbusho International
Science Research Program,\\
Japanese Society for the Promotion of Science (JSPS),\\
German Israeli Bi-national Science Foundation (GIF), \\
Bundesministerium f\"ur Bildung und Forschung, Germany, \\
National Research Council of Canada, \\
Research Corporation, USA,\\
Hungarian Foundation for Scientific Research, OTKA T-029328, 
T023793 and OTKA F-023259.\\

%=======================================================================
%       References
%=======================================================================

%=======================================================================

% likelihood input variables leptonic channel

% likelihood input variables hadronic channel

% likelihood output variables leptonic and hadronic channel

% limits in \kappa_\gamma - \kappa-Z coupling plane

\end{document}